\documentclass[review, times, 12pt]{elsarticle}

\usepackage{hyperref}
\usepackage{graphics}
\usepackage{chemformula}
\usepackage[T1]{fontenc}
\usepackage{dcolumn}
\usepackage{subcaption}
\usepackage{cleveref}
\usepackage{times}
\usepackage{tablefootnote}
\usepackage{siunitx}
\usepackage{multirow}
\usepackage{wasysym} 
\usepackage{xcolor}
\usepackage{makecell}

\usepackage[margin=2.5cm]{geometry}

\usepackage{tabularx}
\usepackage{array}%
\usepackage{tikz}
\usepackage{natbib}

\usetikzlibrary{shapes,arrows}

\newcommand{\figW}{0.961}
\newcommand{\figWH}{0.6}
\newcommand{\figWHH}{0.6}

\newcommand{\spt}{sp$^2$}
\newcommand{\dia}{sp$^3$}
\newcommand{\meth}{CH$_4$}
\newcommand{\hydro}{H$_2$}
\newcommand{\hydrop}{H$_2^+$}
\newcommand{\hydros}{H$_2^*$}

\newcommand{\dpuck}{$\diameter_{\textrm{puck}}$}
\newcommand{\hpuck}{$h_{\textrm{puck}}$}
\newcommand{\imSEM}[1]{\includegraphics[width=0.14\textwidth]{Figures/#1}}
\newcommand{\clamshell}{TM$_{0(n>1)p}$}

\newcommand{\vectE}{{\rm{\bf{E}}}}
\newcommand{\vectH}{{\rm{\bf{H}}}}
\newcommand{\vectB}{{\rm{\bf{B}}}}
\newcommand{\vectEz}{{\rm{\bf{E}}_{\it{z}}}}

\newcommand{\partA}[2]{\frac{\partial#2}{\partial #1}} 
\newcommand{\partB}[2]{\frac{\partial^2#2}{\partial #1^2}} 

\newcommand{\JAC}[1]{{\color{black}#1}}
\newcommand{\JACB}[1]{{\color{black}#1}}

\journal{Journal of \LaTeX\ Templates}

\begin{document}

\begin{frontmatter}
\title{Microwave plasma modelling in clamshell chemical vapour deposition diamond reactors}

\author[CU-PHYSX]{Jerome A. Cuenca\corref{mycorrespondingauthor}}\ead{cuencaj@cardiff.ac.uk}
\author[CU-PHYSX]{Soumen Mandal}
\author[CU-PHYSX]{Evan L. H. Thomas}
\author[CU-PHYSX]{Oliver A. Williams}
\address[CU-PHYSX]{Cardiff School of Physics and Astronomy, Cardiff, Wales, CF24 3AA, UK}
\cortext[mycorrespondingauthor]{Corresponding author}
\begin{abstract}
A microwave plasma model of a chemical vapour deposition (CVD) reactor is presented for understanding spatial heteroepitaxial growth of polycrystalline diamond on Si. This work is based on the \clamshell{} clamshell style reactor (Seki Diamond/ASTEX SDS 6K, Carat  CTS6U, ARDIS-100 style) whereby a simplified \hydro{} plasma model is used to show the radial variation in growth rate over small samples with different sample holders. The model uses several steps: an electromagnetic (EM) eigenfrequency solution, a frequency-transient EM/plasma fluid solution and transient a heat transfer solution at low and high microwave power density. Experimental growths provide model validation with characterisation using Raman spectroscopy and scanning electron microscopy. This work demonstrates that shallow holders result in non-uniform diamond films, with a radial variation akin to the electron density and temperature distribution at the wafer surface. For the same process conditions, greater homogeneity is observed for taller holders, however, if the height is too extreme, the diamond quality reduces. From a modelling perspective, EM solutions are limited but useful for examining electric field focusing at the sample edges, resulting in accelerated diamond growth. For better accuracy, plasma fluid and heat transfer solutions are imperative for modelling spatial growth variation.

\end{abstract}

\begin{keyword}
microwave plasma model, cvd diamond, finite element modelling

\end{keyword}

\end{frontmatter}



\section{Introduction}\label{sec:intro}
Chemical vapour deposition (CVD) has become one of the most popular methods for epitaxial growth of single crystal diamond (SCD) and polycrystalline diamond (PCD) for a wide variety of applications. The basis of this technique requires dissociation of a gaseous hydrocarbon pre-cursor (\meth{}) at low pressures using a reactive \hydro{} species that is excited either using hot filament (HFCVD) \cite{Sachan2019,Narayan2021,Amaral2006,Ali2011,Barber1997,Tabakoya2019,Liang2007} or microwave plasma (MPCVD)\cite{Mandal2021,Tallaire2020,Weng2018,Cuenca2020b,Achatz2006,Benedic2001,Sedov2019,Fendrych2010}. Both HFCVD and MPCVD have their advantages and disadvantages for diamond growth. HFCVD offers easier scalability since the activation region simply depends upon the area covered by the tungsten or tantalum filaments. However, metal incorporation is possible\cite{Ohmagari2018,MehtaMenon1999}, the filament stability is challenging\cite{Okoli1991} and growth rates are moderately low compared to other methods ($\sim$\SI{1.6}{\micro\metre\per\hour}). MPCVD does not have filament issues since the plasma is formed using electromagnetic (EM) standing waves and the growth rates are much higher (>\SI{10}{\micro\metre\per\hour})\cite{Bolshakov2016}, although only over small areas making scalability much harder. This makes MPCVD particularly useful for small, millimetre scale sample growth such as SCD for quantum applications\cite{Achard2020,Mallik2016}.

In MPCVD, understanding the size of the plasma activation region is of huge importance as this directly influences the deposition area and the diamond growth rate. The plasma activation region is affected by several process parameters including forward microwave power, pressure, gas flow rate and temperature in addition to physical parameters such as the sample size, sample holder geometry and of course the reactor topology. The effect of each may be understood empirically or through modelling approaches. Experimental data offers the greatest insight as no CVD reactor is the same as another, especially for bespoke builds. A notable example of this is shown for Asmussen et al. where for SCD, it has been empirically shown that a recessed or `pocket' type sample holder results in less unwanted PCD growth at the sample edges \cite{Nad2016,Wu2016,Charris2017}. Significant material and machining costs are required in order to experimentally iterate geometrical adjustments. Modelling becomes extremely useful at this point and a viable approach for growth optimisation for various reactors, offering faster and cheaper insights into how modified stages, sample holders and reactor walls affect the plasma. Numerous reactor modelling studies exist to this end; the various topologies include the cylindrical TM$_{01p}$ type cavity such as the ASTEX PDS-18, Seki Diamond SDS 5200 series reactors\cite{Funer1999,Gorbachev2001,Shivkumar2016,Silva2010,Yamada2006}, the \clamshell{} type cavity such as the ARDIS-100, Carat Systems CTS6U, Seki Diamond SDS 6K style clamshell \cite{Yamada2012,Yamada2015,Li2014d,Weng2018,Sedov2020a}, the TM$_{02}$ dome style cavity as developed by Su et al.\cite{Su2014} and the ellipsoidal egg-shaped cavity such as the AIXTRON reactor\cite{Funer1999,Li2011a,Li2015}. For a comprehensive review of modelling different microwave reactor topologies, we referred to Silva et al. \cite{Silva2009}.

Plasma modelling is not trivial and requires significant development and experimental validation. Fortunately in this decade a number of commercial packages exist, making this avenue more accessible. One such area which requires more attention from the modelling perspective is the sample holder design. Shivkumar et al. have contributed significant understanding of pillar type models to focus the plasma, corroborated with optical emission spectroscopy\cite{Shivkumar2016}. A notably recent study by Sedov et al. combines both modelling and experimental growths of the geometrical effect of recessed and pedestal type sample holders for 2" Si wafers using an E-field model, demonstrating that pedestal holders yielded higher quality diamond films when compared to a recessed holder\cite{Sedov2020}.

In this work, we demonstrate a simple microwave plasma model of the \clamshell{} reactor (Seki Diamond 6K style) that can be implemented in COMSOL Multiphysics\textregistered{} for the purpose of sample holder or `puck' design. The model presented here uses a simplified \hydro{} reaction cross-section set currently accessible from the Itikawa database on lxcat.net \cite{Itikawa2009}. A simple experimental validation is achieved using a sample puck of varying height, positioning the sample closer or further away from the plasma. The model is compared with experimental characterisation of thin film diamond growths over small Si wafers ($\diameter=$ 1", $t=$ \SI{0.5}{\milli\metre}). In Section \ref{sec:theory} the EM theory is briefly discussed for \clamshell{} style reactors along with the plasma and heat transfer continuity equations for the finite element model (FEM). In Section \ref{sec:model} the modelling implementation is shown, including the boundary conditions and the modelling results for. In Section \ref{sec:exp} the experimental data is presented, including plasma images from the viewports, Raman spectra and scanning electron microscopy (SEM) images of the films.

\section{Theory}\label{sec:theory}
\subsection{Electromagnetic field}
The microwave plasma is sustained by the electric (E) field which accelerates the seed electrons to interact with the \hydro{} gas molecules. The shape and location of the microwave plasma activation region is therefore dependent upon the spatial EM field within the resonant chamber. One of the quickest methods of modelling the plasma location is to simply calculate the EM field distribution of the resonant mode through eigenfrequency analysis\cite{Silva2010}. Analytically, these standing wave distributions are determined by deriving the Helmholtz resonator solution from the time-harmonic Maxwell’s equations:
\begin{align}
	\nabla\cdot\vectE = & \rho_c / \varepsilon_0 \label{eq:max1} \\
	\nabla\cdot\vectB = & 0 \label{eq:max2} \\
	\nabla\times\vectE = & -j\mu_0\mu_r\omega \rm{\bf{H}} \label{eq:max3}\\
	\nabla\times\vectH = & j\varepsilon_0\omega\varepsilon_r \vectE \label{eq:max4}
\end{align}
where $\vectE$ and $\vectH$ are the electric and magnetic fields, respectively, $\varepsilon_r$ and $\mu_r$ are the relative permittivity and permeability of the medium, respectively, $\rho_c$ is the charge density, $\varepsilon_0$ and $\mu_0$ are the permittivity and permeability of vacuum, respectively, $\vectB=\mu_r \mu_0 \vectH$ and $\omega$ is the angular frequency. A Helmholtz resonator solution is obtained using vector identities. For transverse magnetic (TM) modes where $\vectH_{\it{z}}=0$ and $\vectEz$ is finite: 
\begin{equation}
	\nabla^2\vectEz -k^2\vectEz = 0
\end{equation}
where $k^2=\omega^2\varepsilon_0\varepsilon_r\mu_0\mu_r$ is defined as the wavenumber. Analytical solutions can be derived based on the coordinate system, or solved for using FEM.

Microwave reactor topologies are typically cylindrical or elliptical since the standing wave patterns are based upon Bessel functions which inherently focus the $E$-field, and therefore the plasma, into the central regions of the cavity\cite{Silva2010}. For example in cylindrical coordinates, $\nabla^2\vectEz$ becomes a Poisson equation:
\begin{equation}
	\partB{r}{\vectEz} + \frac{1}{r}\partA{r}{\vectEz} +\frac{1}{r^2} \partB{\theta}{\vectEz} + \partB{z}{\vectEz} -k^2 \vectEz = 0
\end{equation}
The radial component has solutions dependent on Bessel functions and the azimuthal and axial components have solutions based on sinusoidal functions. By imposing boundary conditions, the cylindrical TM $\vectEz$ field can be obtained:
\begin{equation}
	\vectEz(r,\theta,z) = J_n\left(\frac{\alpha_{mn}}{a}r\right) \textrm{ cos}\left(n\pi\theta\right) \textrm{ cos}\left(\frac{p\pi}{l}z\right)
	\label{eq-TMmain}
\end{equation}
where $J_n$ is the $n^{\rm{th}}$ order Bessel function $a$ and $l$ are the radius and height of the cylinder, respectively, $\alpha_{mn}$ is the $n^{\rm{th}}$ troot of the $m^{\rm{th}}$ Bessel function and $p$ is the integer number of axial standing waves.
One can then derive all other components of the field distribution using (\ref{eq:max1}) to (\ref{eq:max4}), however, from this equation it is clear which TM modes will be useful for an MPCVD reactor. From a practical point of view, the substrate should be placed in the centre of the reactor as to be as far from the walls as possible to avoid any etch or re-deposition of wall contaminants. Firstly, this means that $p>0$ otherwise the absence of an axial standing wave would directly connect the plasma to the top and bottom of the cylindrical cavity walls. Secondly, in order to achieve a centralised plasma, only TM modes which consider E-fields where $m=0$ can be used as this is the only Bessel function where the result is non-zero and finite at $r=0$. It has been shown that modes where $m>0$ can actually be used to monitor the temperature of a cavity resonator as they are less sensitive to the centre\cite{Cuenca2017,Cuenca2017a}. Bessel functions roughly decay proportional to $r^{-1/2}$ and with increasing root and frequency, the E-field is further compressed and concentrated into the centre of the cavity. Thus, the ideal case is to use the highest $n$ possible although this would make a cavity with a very large radius. This defines the conventional use of cylindrical reactors based on TM$_{0np}$ modes as it places the E-field central and at either the top or bottom of the cavity. A sample holder becomes crucial to break this degeneracy by disrupting the E-field and encourages the plasma to be localised to only one of these regions.

While the E-field distribution provides a general idea as to where the plasma is to be localised, the simple EM field eigenfrequency approximation does not take into account the perturbation of the diffuse but conductive gas. The relative permittivity of the conductive gas is complex and can be modelled using the Drude-Lorentz model:
\begin{align}
	\varepsilon_r = 1 - \frac{\omega_p^2}{\omega^2+\nu_m^2} - j\left( \frac{\omega_p^2\nu_m}{\omega(\omega^2+\nu_m^2)}      \right)\\
	\omega_p = \sqrt{\frac{e^2n_e}{\varepsilon_0m_e}}
\end{align}
where $\omega_p$ is the plasma frequency and $\nu_m$ is the electron-species collision frequency, $n_c$ is the electron density and $e$ and $m_e$ are the electron charge and mass, respectively. To introduce the metallic gas of a certain electron density, the microwave frequency and information on how frequently the electrons interact with the pre-cursor gas is needed. Notable methods to incorporate this are the early F\"{u}ner models, where at a defined threshold E-field value, $n_e$ would be finite and zero otherwise\cite{Shivkumar2016,Funer1999,Funer1995}. While this method is suitable at high pressure such that the plasma ball is confined within the E-field region, this never allows the plasma to be situated in the nodal regions of the standing wave, which contradict the large pancake plasma shapes typically found in \clamshell{} style reactors at low pressures. 

\subsection{Plasma fluid}
\begin{figure}[t!]
  \centering
  \includegraphics[width=\figWH\textwidth]{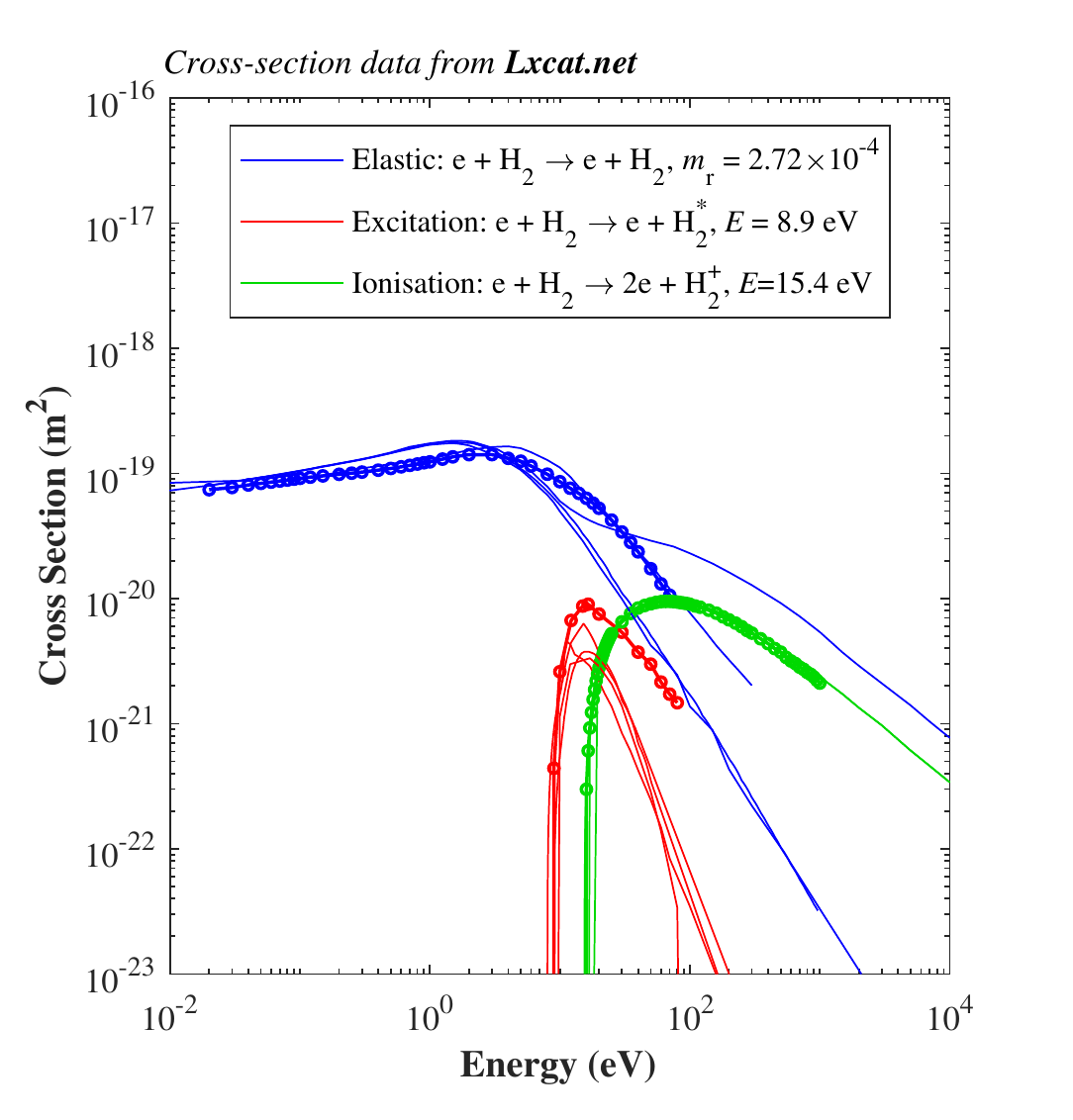}
  \caption{Simplified electron impact cross section reactions with \hydro{} from the Biagi (MAGBOLTZ), CCC\cite{Bray1992,Zammit2014,Zammit2016}, IST Lisbon, Itikawa\cite{Itikawa2009} and Phelps\cite{Buckman1985} databases (www.lxcat.net, retrieved on September 3$^{\rm{rd}}$, 2021). Data used in this study (denoted '$\circ$' ) is from the Itikawa database with parameters given in the legend.}
  \label{fig-xsec}
\end{figure}

To fully incorporate the nuance of pressure in larger area reactors, the fluid model is introduced which allows modelling of the collective behaviour of the electrons, ions and neutral species\cite{Yamada2011,Yamada2007,Yamada2006,Hassouni1999}. These gaseous species are initially distributed homogeneously within the cavity and several energy dependent electron-impact reactions are defined with a rate constant or reaction cross section. An example of a simplified reaction set is shown in Fig. \ref{fig-xsec}. The accelerated electrons may result in elastic scattering (e + \hydro{} $\rightarrow$ e + \hydro{}), excitation of a species (e + \hydro{} $\rightarrow$ e + \hydros{}), ionisation (e + \hydro{} $\rightarrow$ e + \hydrop{}) or attachment or detachment of atoms (e + \hydro{} $\rightarrow$ e + 2H). Additional reactions can also occur with these products, producing a soup of various charged and neutral species. As electrons and ions are produced, electrostatic forces and concentration gradients then result in a plasma fluid diffusing around the high E-field regions. The key advantage of the fluid approach is that the plasma has a finite density, allowing the fluid to be modelled as a function of pressure, temperature and even the gas flow velocity. 

For the electrons, the fluid is modelled using continuity equations of motion and energy conservation equations. The equation of motion for the number density is:
\begin{align}
	\partA{t}{n_e} + \nabla\cdot{\bf\Gamma}_e = R_e \label {eq:cont1a}\\
	{\bf{\Gamma}}_e = -\mu_e n_e \bf{E}_{\rm{a}} - \nabla \it D_e n_e 
	\label{eq:cont1b}
\end{align}
\JACB{where $n_e$ is the electron density, ${\bf\Gamma}_e$ is the electron flux vector, $\mu_e$ is the electron moblity, $D_e$ is the electron diffusivity, ${\bf{E}}_a$ is the ambipolar field and $R_e$ represents the electrons that are either produced or consumed during impact reactions. The first term in ${\bf\Gamma}_e$ is associated with the ambipolar E-field and is the E-field that is generated by the separation of the ions and the electrons.} The second term is associated with drift contributions from concentration gradients. Coupled with (\ref{eq:cont1a}) is an electron energy conservation equation of a similar form. Energy is either lost or gained from elastic/inelastic reactions with the gaseous species, absorbed from microwave heating in the E-field or accelerated in the electrostatic ambipolar fields:
\begin{align}
	\partA{t}{n_{\varepsilon}} + \nabla\cdot{\bf{\Gamma}}_{\varepsilon} +{\bf{E}}_{\rm{a}}\cdot{\bf{\Gamma}}_e = S_{\varepsilon} +\frac{Q_{\rm{mw}}}{e} \\
	\Gamma_{\varepsilon} = -\mu_{\varepsilon} n_{\varepsilon} \bf{E}_{\rm{a}} - \nabla \it D_{\varepsilon} n_{\varepsilon}
	\label{eq:cont2}
\end{align}
\JACB{where $n_{\varepsilon}$ is the electron energy density, ${\bf{\Gamma}}_\varepsilon$ is the electron energy flux vector, $\mu_\varepsilon$ is the electron energy mobilty, $D_\varepsilon$ is the electron energy diffusivity, $S_\varepsilon$ is the energy gain or loss from impact reactions and $Q_{\rm{mw}}$ is the microwave heating of the electrons.}

For the heavier gas species such as ions and neutral molecules, the continuity equations are similar to the electrons, however, may include inertial terms from the background gas flow velocity (omitted in this model). For multiple reaction species (\hydro{}, \hydrop{} and \hydros{}), the continuity relation for the $i^{\rm{th}}$ specie is:
\begin{align}
	\rho_i\partA{t}{w_i} = \nabla\cdot{\bf{\Gamma}}_i + R_i \\
	{\bf{\Gamma}}_i = \rho_i w_i v_d
	\label{eq:cont3}
\end{align}
where $\rho_i$ is the density, $w_i$ is the mass fraction, ${\bf\Gamma}_i$ is the ion flux vector, $R_i$ represents the ions that are either produced or consumed in reactions and $v_d$ is the \JAC{species diffusion velocity}.

\subsection{Heat transfer model}
In addition to the plasma fluid model, the gas temperature (or neutral/ion species temperature) is calculated over time assuming a simple conductive heat transfer model using a mass averaged gas density, heat capacity, thermal conductivity and the EM power dissipated. The continuity equations for this calculation are:
\begin{align}
	\rho_n C_p \partA{t}{T}= \nabla\cdot(k\nabla T)  + Q_{\rm{mw}}
	\label{eq:cont4}
\end{align}
where $\rho_n=pM_n/RT$ is the gas density, $p$ is the pressure, $M_n$ is the mean molar mass, $R$ is the gas constant, $T$ is the temperature, $C_p$ is the heat capacity and $k$ is the thermal conductivity.

\section{Modelling}\label{sec:model}
\subsection{Method}

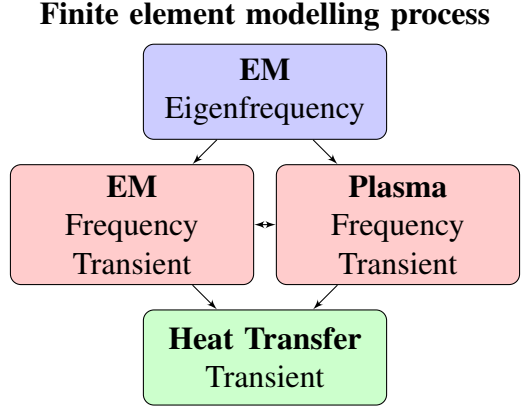
\begin{figure}[t!]
  \centering
\tikzstyle{block0} = [rectangle, fill=none, 
    text width=16em, text centered, rounded corners, minimum height=2em]
\tikzstyle{block1} = [rectangle, draw, fill=blue!20, 
    text width=7em, text centered, rounded corners, minimum height=3em]
\tikzstyle{block2} = [rectangle, draw, fill=red!20, 
    text width=7em, text centered, rounded corners, minimum height=3em]
\tikzstyle{block3} = [rectangle, draw, fill=green!20, 
    text width=7em, text centered, rounded corners, minimum height=3em]
\tikzstyle{line} = [draw, -latex']
\tikzstyle{dline} = [draw, latex'-latex']

\begin{tikzpicture}[node distance = 2cm, auto]
  \renewcommand{\baselinestretch}{1} 
    \node [block0] (title) {\textbf{Finite element modelling process}};
    \node [block1, below of=title, node distance=1cm] (eme) {\textbf{EM} \\ Eigenfrequency};
    \node [block2, below left of=eme, node distance=2.5cm] (emf) {\textbf{EM} \\ Frequency \\ Transient};
    \node [block2, below right of=eme, node distance=2.5cm] (plas) {\textbf{Plasma} \\ Frequency \\ Transient};
    \node [block3, below right of=emf, node distance=2.5cm] (ht) {\textbf{Heat Transfer}  \\ Transient};

    \path [line] (eme) -- (emf);
    \path [line] (eme) -- (plas);
    \path [dline] (emf) -- (plas);
    \path [line] (emf) -- (ht);
    \path [line] (plas) -- (ht);
\end{tikzpicture}
  \caption{Finite element modelling process flow using COMSOL Multiphysics\textregistered.}
    \label{fig-flow}
\end{figure}

The FEM process is split into three separate studies as shown in Fig. \ref{fig-flow}. First, the EM model is run where the eigenfrequencies of the geometry are calculated to ensure that the correct mode is identified (the \clamshell{} type mode where a TM$_{011}$ distribution is present in the active region of the reactor). This step is crucial for determining if any reactor modifications or the introduced sample holders shift the frequency away from the source generator frequency range. Additionally, higher order modes which are not necessary for diamond growth can also be identified. Secondly, the frequency-transient electromagnetic/plasma model is calculated at the eigenfrequency. The reactor port power is varied and provides a  continuous wave to set up the EM standing wave. In this way the E-field intensity, and therefore the plasma, is power dependent. The electrons and gaseous species are distributed within the cavity and the transient response is modelled from 0 to 10 s to allow the plasma fluid to evolve to a steady state at an initial ignition power and pressure (1.5 kW at 20 mbar) or low microwave power density (MWPD). Subsequently, the MWPD is \JACB{ramped up} to growth conditions or high MWPD (5 kW at 160 mbar) over the simulated time of 1 hour to keep the solution stable. Finally, the third step calculates a heat transfer solution to obtain the gas temperature using the microwave power dissipated in the plasma. 

\begin{figure}[t!]
  \centering
  \includegraphics[width=\figWH\textwidth]{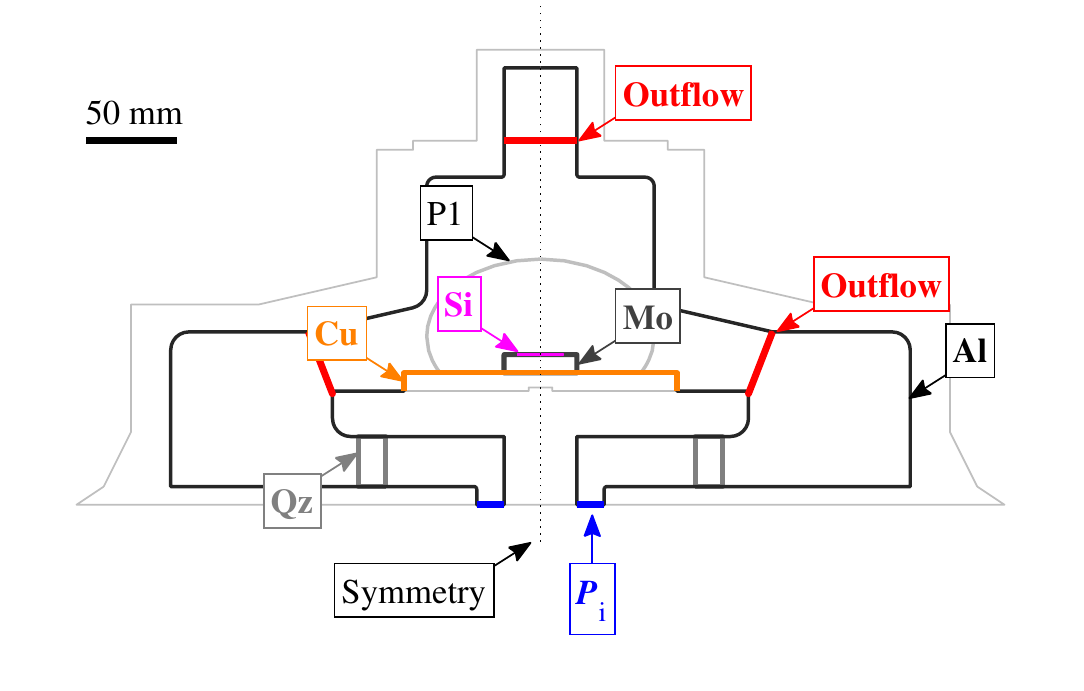}
  \caption{Simplified schematic of the clamshell reactor (Carat Systems CTS6U). Wall surfaces are modelled as impedance boundary conditions. The dotted line denotes axial symmetry about $r=0$. $P_i$  denotes a lumped coaxial port boundary condition. The red outflow lines denote the limits of the fluid simulation as electron and heat outlets for the plasma and heat transfer solutions, respectively. P1 denotes the location of the plasma region.}
  \label{fig-model}
\end{figure}

The cavity boundary conditions are also shown in Fig. \ref{fig-model}. The mesh is kept consistent in all steps, utilising a free quadratic with boundary layers at the extremities for the plasma solution. The mesh also forces a distribution of 50 nodes across the Si sample surface to ensure a high resolution for the spatially dependent electron densities. For the first EM eigenfrequency model, the chamber domain is assumed vacuum and the walls are all modelled as metallic impedance boundary conditions ($\sigma>10^7$ S/m). To reduce computation time in the plasma model, the domain is only confined to the centre of the reactor, marked by the outflow/electron outlet conditions in Fig. \ref{fig-model}. The gas pressure of this domain is varied from 20 to 160 mbar. For the plasma model, all walls are defined as grounds with additional surface reactions for excited species to relax to neutral species (\hydros{} $\rightarrow$ \hydro{}, \hydrop{} $\rightarrow$ \hydro{}). The cross-section reactions for the gaseous species have adopted a simplified hydrogen plasma similar to Yamada et al. \cite{Yamada2006} to reduce computation time with a dataset obtained from the Itikawa database\cite{Itikawa2009} (available on lxcat.net). The reactor port $P_i$ is defined as a lumped port with excitation varying from 1.5 to 5 kW. For the heat transfer model, heat flux boundary conditions are imposed to the extremities to ensure that the sample holder stage is at the correct temperature of approximately \SI{800}{\celsius}. The heat flux boundaries simulate the cooling of the reactor using a fixed systematic heat transfer coefficient for all sample holder heights (a modelled cooling flux of \SI{850}{\watt\per\metre\kelvin} for all external boundaries with an ambient temperature of \SI{25}{\celsius}). The sample is modelled as a Si wafer ($\diameter=$1", $t=$ \SI{0.5}{\milli\metre}) and is positioned on top of the Mo sample holder puck ($\diameter_{\rm{puck}}=40$ to \SI{60}{\milli\metre}, $h_{\rm{puck}}=$ 1 to \SI{20}{\milli\metre}). The corners of the sample are rounded with a radius of \SI{0.1}{\milli\metre} and the Mo holder rounded with a radius of \SI{0.2}{\milli\metre} to ensure a high mesh density at the anticipated high E-field regions and avoid convergence errors.

 \subsection{Electromagnetic model}
 
\begin{figure}[t!]
  \centering
  \includegraphics[width=\figWH\textwidth]{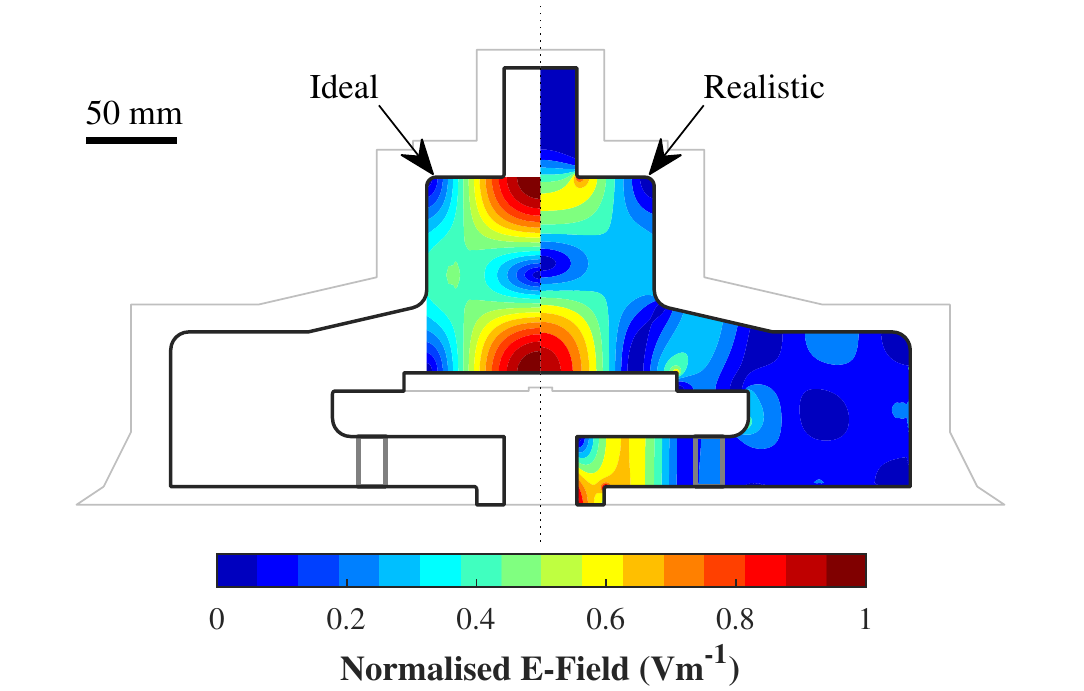}
  \caption{The base EM eigenfrequency model without the Mo sample holder puck. The E-field is normalised to the maximum value. For comparison and identification of the correct mode, the ideal cylindrical case of the TM$_{011}$ mode (left side) and the realistic \clamshell{} design (right side) are shown.}
  \label{fig-em}
\end{figure}

\begin{figure}[t!]
  \centering
  \includegraphics[width=\figWH\textwidth]{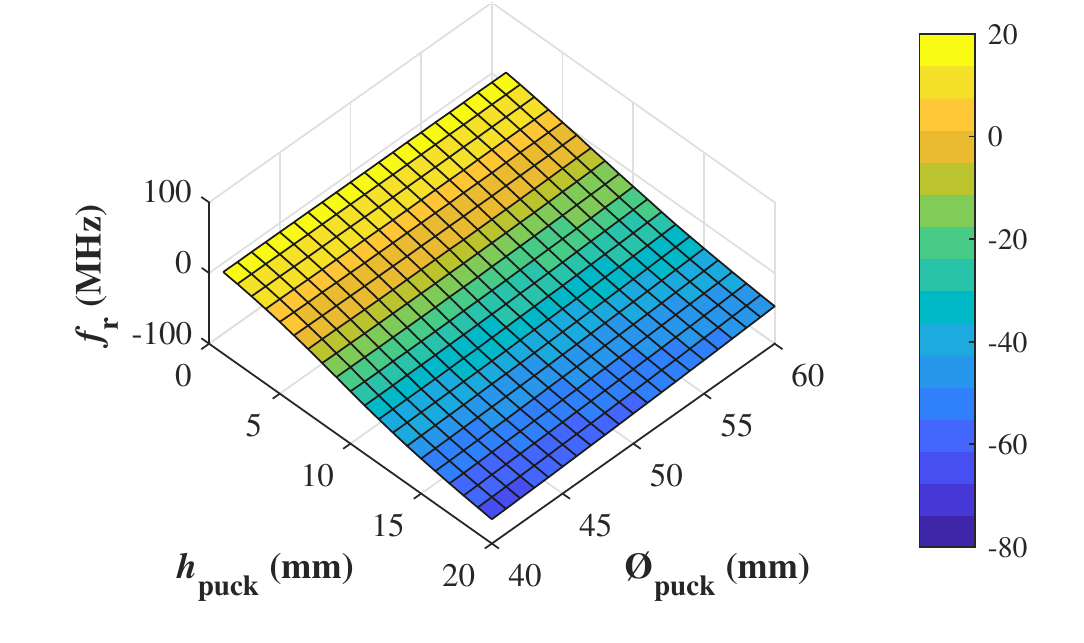}
  \caption{Modelled shift in resonant frequency caused by the Mo sample holder puck of varying dimensions; \hpuck{} and \dpuck{} denote the height and diameter, respectively. The sample holder has a Si wafer sample on top ($\diameter=$ 1", $t=$ \SI{0.5}{\milli\metre}). }
    \label{fig-freq}
\end{figure}

Figure \ref{fig-em} shows the EM eigenfrequency model with the E-field distribution of an ideal TM$_{011}$ mode and the \clamshell{} reactor without the Mo sample puck. The \clamshell{} reactor mode shows a reasonably similar E-field distribution to the ideal case, demonstrating that the correct mode has been identified. In this mode, there is an E-field node separating the central region and a side lobe. High intensity E-field regions at the edge of the stage are found where a secondary plasma can also be sustained. Figure \ref{fig-freq} shows how the calculated resonant frequency of the EM model is perturbed to lower frequencies as a Mo sample puck is introduced. With the current dimensions, the initial unperturbed resonant frequency is calculated at approximately 2.45 GHz \JACB{with a -3 dB bandwidth of $\sim$\SI{87}{\mega\hertz}}. The results show that wider pucks have less of a frequency shift, while taller pucks alter the resonant frequency by as much as $\Delta f\approx$ -66 MHz at \hpuck{} = 20 mm. This demonstrates that shorter sample pucks are more favourable for stable operation with a magnetron with a fixed frequency output. \JACB{Interestingly, this is not so much a problem for solid state sources since the signal generator frequency can varied.} 

After introducing the puck, the E-field distribution of the EM model is shown in Fig. \ref{fig-main}(a). The E-field is perturbed and high intensity regions occur at the corners of the puck. This is simply because of Maxwell's equations (\ref{eq:max1}) and (\ref{eq-TMmain}); the $E_z$ field lines should be perpendicular to the Cu stage and the introduction of a metal object creates parallel surfaces that disrupts this condition resulting in a reconfiguration of surface currents. With increasing sample holder puck height, the E-field, and therefore the plasma activation region, is concentrated towards the edges of the holder. The nodal regions of the E-field are also clearly visible either side of the puck where, based on the threshold modelling approach, the plasma could not exist; or if it did, the plasma could exist in multiple regions in the cavity. It is also noted that with increasing sample holder height, the E-field hot spots at the top of the chamber reduce in intensity, decreasing the likelihood for a secondary plasma to ignite at these regions. 

 \subsection{Plasma Model}
\begin{figure*}[t!]
  \centering
  \includegraphics[width=\figW\textwidth]{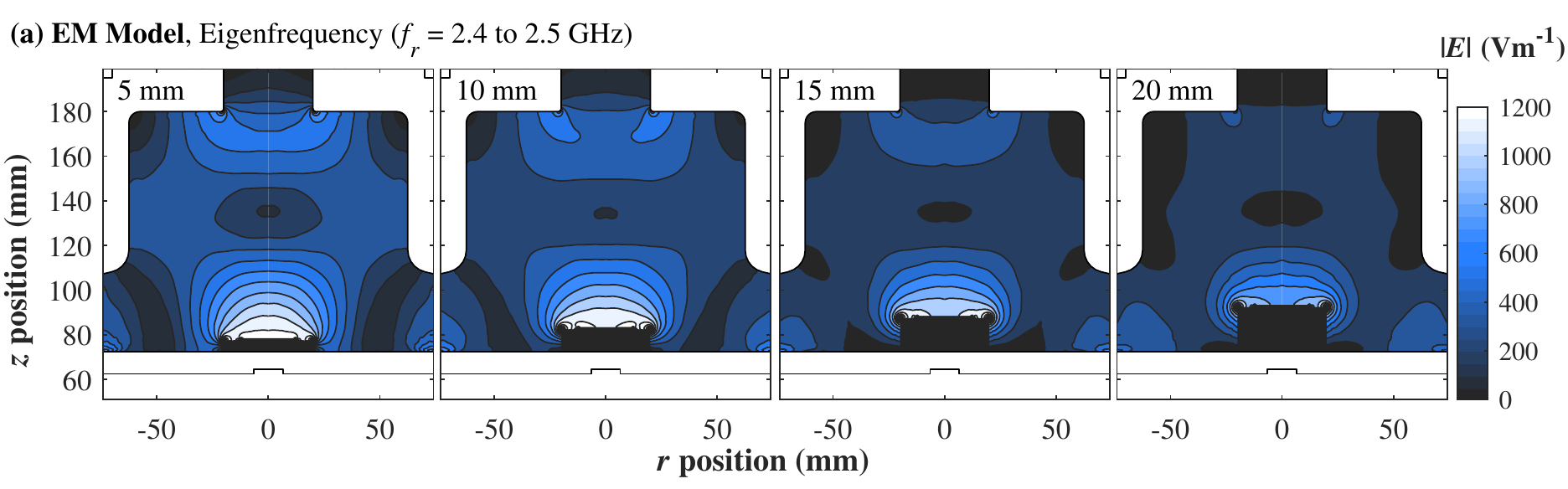}
  \includegraphics[width=\figW\textwidth]{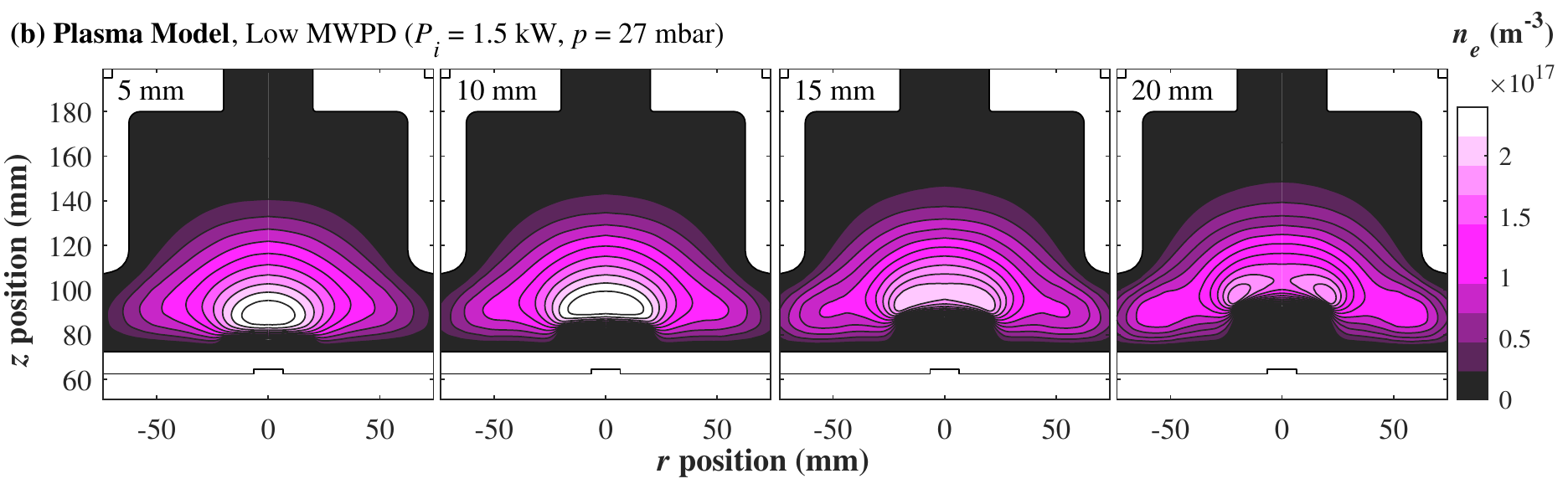}
  \includegraphics[width=\figW\textwidth]{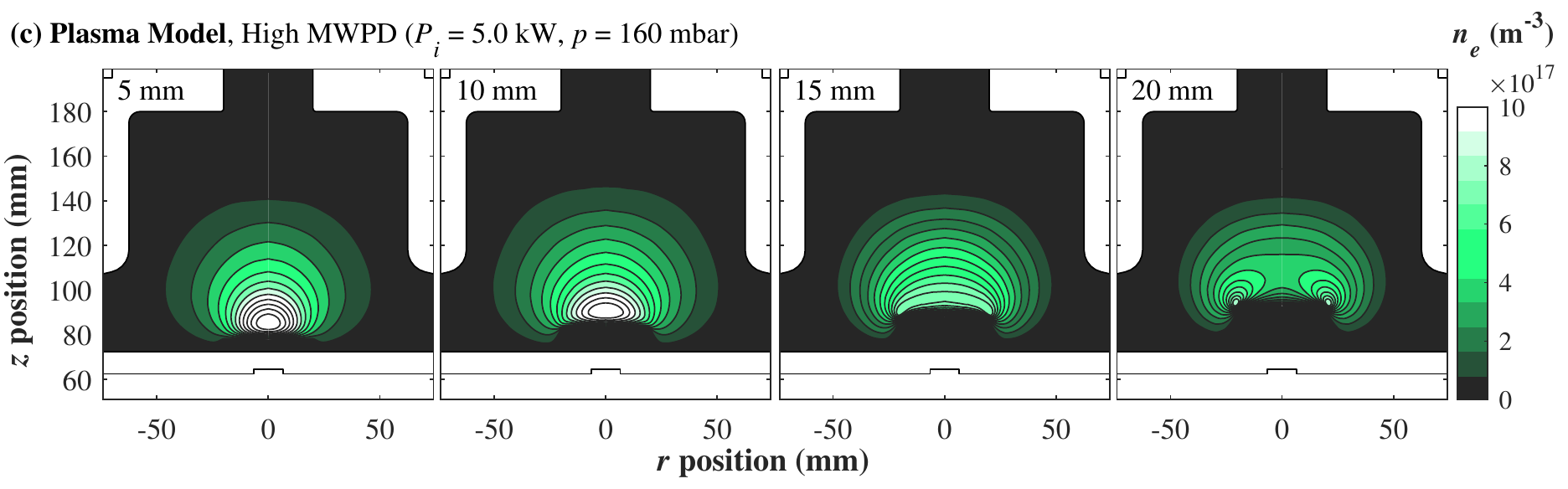}
  \includegraphics[width=\figW\textwidth]{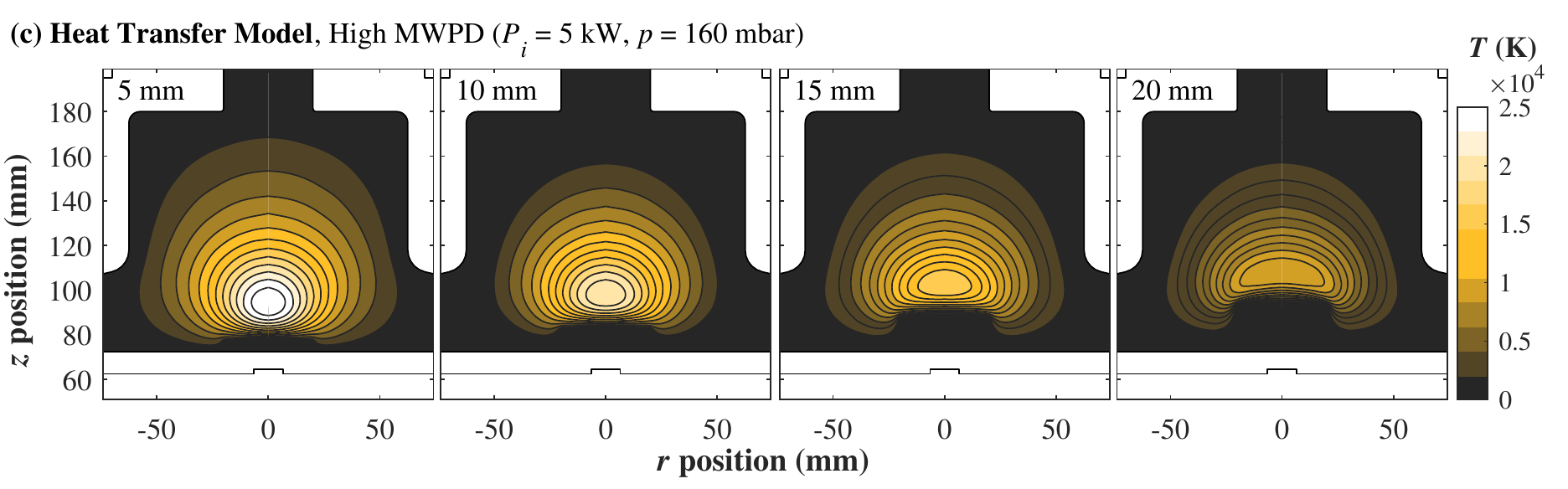}
  \caption{Microwave hydrogen plasma model using the simplified eigenfrequency, plasma coupled and heat transfer solution approach for varying sample holder heights (\hpuck{} = 5 to 20 mm) with a Si wafer on top ($\diameter=$ 1", $t=$ \SI{0.5}{\milli\metre}). Model uses cross-section data from the Itikawa database, available on lxcat.net\cite{Itikawa2009}.}
    \label{fig-main}
\end{figure*}

The plasma fluid models are shown in Fig. \ref{fig-main}(b) and \ref{fig-main}(c) and clearly demonstrate a focused electron density in the sample region. The calculated electron densities are as high as \SI{2.2\times10^{17}}{\per\metre\cubed} and \SI{10\times10^{17}}{\per\metre\cubed} in the low and high MWPD models, respectively. Based on a $\omega_p=$ \SI{2.45}{\giga\hertz}, $n_c =$ \SI{7.45 \times 10^{16}}{\per\metre\cubed}, stipulating that at these electron densities, the microwaves are not able to propagate freely through the plasma and thus attenuates, thereby depositing microwave power into the plasma\cite{Silva2009}. Note that in the low MWPD solution the electron density distribution is much wider than the E-field result, demonstrating that simple EM solutions are potentially less appropriate for modelling low pressure plasmas. \JACB{Although PCD diamond growth with low non-diamond carbon impurities typically occurs at high MWPD, the lower pressure solutions are an imperative result for several purposes. The first is that some applications involve nano-crystalline diamond (NCD) and ultra-nanocrystalline diamond (UNCD) which is typically grown at lower power densities\cite{Cuenca2019,Williams2007,Sankaran2018} as well as hybrid graphene-diamond films\cite{Carvalho2016}. The second is to ensure that the plasma can actually be ignited at the right place in the chamber.} Finally, investigating large area growth using lower pressure would be challenging using an EM solution alone. At low MWPD and at large heights of 15 to 20 mm, the sample is pushed further into the the plasma and the fluid moves towards the electron outlets at the side of the stage. This is not favourable as the risk of the plasma pushing to below the stage towards the quartz ring region where the microwaves enter is much greater. At high MWPD, the plasma becomes the familiar elliptical shape situated over the sample with a smaller area and a much higher electron density (\SI{n_e\sim10\times10^{17}}{\per\metre\cubed}), similar electron densities to those found in previous models of different reactors at growth conditions\cite{Yamada2007,Kelly2012,Hassouni1999}. At high MWPD, higher Mo sample pucks result in the plasma further focusing towards the edges which will inherently affect the spatial CVD diamond growth rate across the sample.

 \subsection{Heat Transfer Model}

\begin{figure}[t!]
  \centering
  \includegraphics[width=\figWH\textwidth]{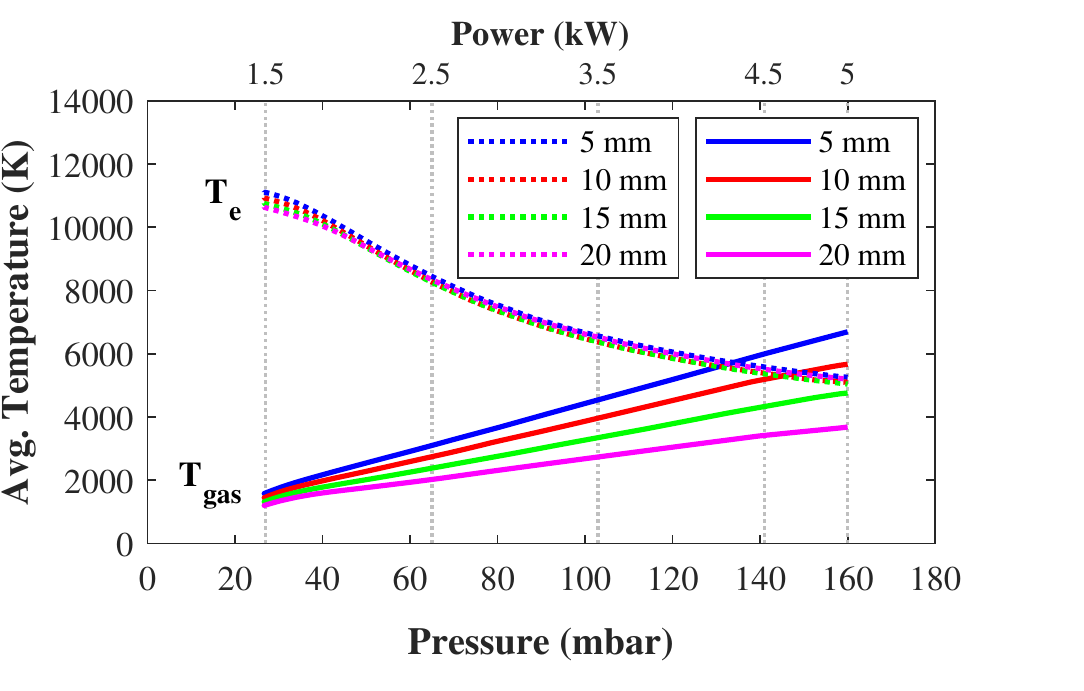}
  \caption{Average gas temperature (solid) and electron temperature (dotted) over the plasma fluid domain region (P1 in Fig. \ref{fig-model}) at different sample holder heights.}
    \label{fig-TevsT}
\end{figure}

The heat transfer solution at high MWPD shows that the temperature of the gas reaches several tens of thousands Kelvin. Although these values are much higher than those reported by Shivkumar et al. in cylindrical TM$_{01p}$ type reactors ($\sim$2,500 K) \cite{Shivkumar2016}, in that study the modelled MWPD was lower (700 W at 30 Torr) which cannot be easily sustained in the \clamshell{} reactor. The gas temperature is hotter at shallower puck heights and decreases significantly as the puck is pushed into the plasma. This is likely due to the fact that the sample puck is in direct contact with the heavily cooled Cu stage underneath and an increasing volume of Mo increases the thermal mass of the puck, thereby reducing the temperature. Figure \ref{fig-TevsT} shows the averaged gas and electron temperatures over the ellipse drawn over the plasma region (defined as P1 in Fig. \ref{fig-model}). Here, it becomes clear that at low MWPD, the plasma is not at thermodynamic equilibrium as the electron temperature ($T_e$) is much higher than the background gas temperature ($T_g$) at low MWPD.  Increasing the gas pressure increases the number of electron-hydrogen collisions which increases $T_g$ and reduces $T_e$.
 At high MWPD, for a modelled central substrate temperature of approximately \SI{800}{\celsius}, the plasma tends towards a collisional plasma condition. 

\section{Experiment\label{sec:exp}}
\subsection{Experimental method}
To demonstrate the overall affect of the sample holder height on the CVD diamond process, three Mo pucks were machined ($\diameter=$ \SI{40}{\milli\metre}, $h_{\rm{puck}}=$ 5, 10 and \SI{15}{\milli\metre}) and used for thin film growths (approximately \SI{1}{\micro\metre} thick) on small Si wafers ($\diameter=$ 1'', $t=$ \SI{0.5}{\milli\metre}). It is worth stipulating that a fourth puck ($h_{\rm{puck}} = 20$ mm) was also machined, however, stable plasma ignition was not possible. This is likely due to the holder significantly perturbing the resonant frequency of the chamber. The Si wafers were seeded using the ultrasonic seeding process\cite{Williams2011, Mandal2021a}. Briefly, the wafers were solvent cleaned and placed in a nanodiamond colloidal solution with a particles of a positive zeta potential whilst under ultrasonic agitation for 10 minutes. The samples were then rinsed in deionised water, dried using an air gun and placed on the top of the sample holder inside \JACB{ of a \clamshell{} style reactor for CVD diamond growth}. Samples were grown using a \meth{}/\hydro{} gas mixture with a \meth{} concentration of 3\% in a total flow rate of 300 sccm at a forward microwave power of 5 kW at 160 mbar. The total growth time for all samples was fixed at 30 minutes, followed by a 5 minute cool down ramp to 1.5 kW at 27 mbar. The temperature at growth MWPD was monitored using a Williamson dual wavelength pyrometer (DWF-24-36C), giving initial measured readings of 760, 790 and \SI{780}{\celsius} for the 5, 10 and 15 mm growths, respectively. Two samples for each molybdenum puck height were grown at different reactor usage times.

After growth, the samples were examined using Raman spectroscopy and scanning electron microscopy (SEM). Raman spectroscopy was conducted using a Horiba LabRAM HR Evolution with a green laser ($\lambda=$ \SI{532}{\nano\metre}) and a grating of \SI{600}{l\per\milli\metre} from 200 to \SI{2000 }{\centi\metre}$^{-1}$ to allow for sensitivity to both the diamond and non-diamond carbon content. Line scans were taken at points across the samples ($N$ = 20 points over \SI{22}{\milli\metre}). SEM images were obtained using a Hitachi SU8200 (\SI{10}{\kilo\volt} at \SI{10}{\micro\ampere}) with a working distance of \SI{8}{\milli\metre}.

\subsection{Microwave Plasma}
\begin{figure}[t!]
  \centering
  \includegraphics[height=2.5cm]{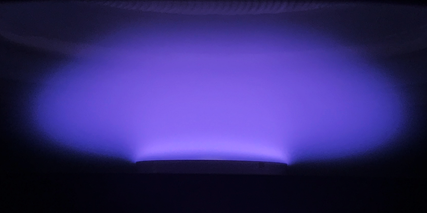} \includegraphics[height=2.5cm]{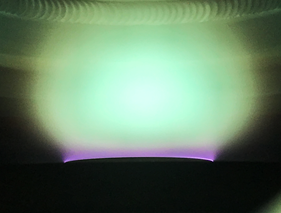}\\
  \includegraphics[height=2.5cm]{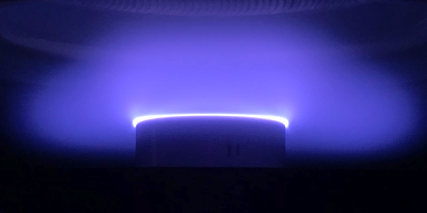} \includegraphics[height=2.5cm]{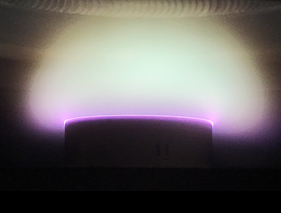}
  \caption{Photographs of the microwave \hydro{}/\meth{} plasma at low (left column) and high (right column) MWPD for sample puck heights of 5 mm (top row) and 15 mm (bottom row), respectively. Low MWPD plasma is at 1.5 kW at 27 mbar, while high MWPD is 5 kW at 160 mbar with 3\% \meth{} in a total flow rate of 300 sccm. Images were taken from the side viewport of the Carat Systems CTS6U using an Apple iPhone 7.}
  \label{fig:photo}
\end{figure}

Figure \ref{fig:photo} shows the typical images of a microwave \hydro{}/\meth{} plasma at ignition and at the point of reaching diamond growth conditions with different puck heights in the \clamshell{} reactor. At low power density the plasma emits a purple glow associated with the combined emissions of the H$_\alpha$ ($\sim$657 nm), H$_\beta$ ($\sim$486 nm) and the H$_\gamma$ ($\sim$437 nm) lines\cite{Hemawan2015}. As the MWPD is increased, the plasma becomes the familiar green ellipsoid over the sample holder, characteristic of C$_2$ emission from the small \meth{} concentration. The images show that the low pressure plasma extends well beyond the extents of the puck, into where the EM E-Field nodes are calculated to be. The edges of the 15 mm puck are also much brighter compared to the 5 mm puck owing to the E-field focusing. The increase in MWPD results in a smaller and focused ellipsoid above the puck, with the higher 15 mm puck distorting the bottom edges shape of the plasma.

\begin{figure*}[t!]
\setlength{\tabcolsep}{2pt}
\renewcommand{\arraystretch}{1}
	\centering
		\begin{tabular}{>{\centering}m{0.5in} c c c c c c}
		{} & \bf{centre} & & & & & \bf{edge}\\
		{} & 0 mm & 2.5 mmm & 5 mm & 7.5 mm & 10 mm & 12.5 mm \\
		5 mm	&	\imSEM{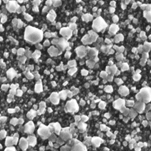}	&	\imSEM{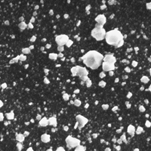}	&	\imSEM{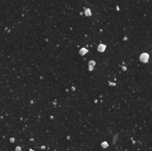}	&	\imSEM{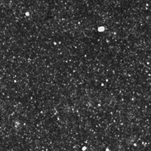}	&	\imSEM{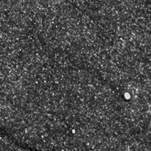}	&	\imSEM{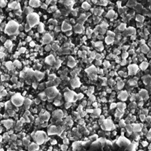}	\\
		10 mm &	\imSEM{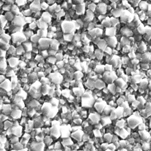}	&	\imSEM{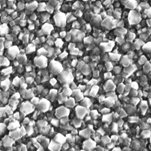}	&	\imSEM{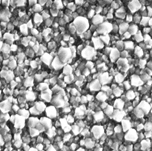}	&	\imSEM{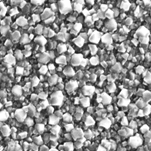}	&	\imSEM{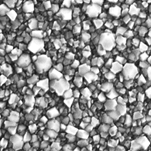}	&	\imSEM{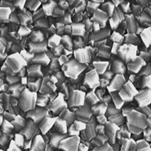}	\\
		15 mm &	\imSEM{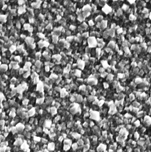}	&	\imSEM{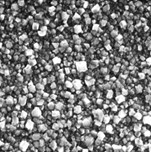}	&	\imSEM{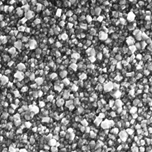}	&	\imSEM{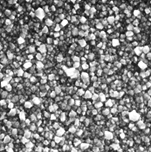}	&	\imSEM{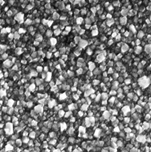}	&	\imSEM{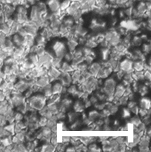}	\\
		\end{tabular}

  \caption{SEM images of the CVD diamond films grown at varying Mo sample holder heights (5, 10 and \SI{15}{\milli\metre}). Images are taken at fixed radial distances from the centre of the wafer to the edge. Scale bar represents a \SI{5}{\micro\metre} length.}
  \label{fig-sem}
\end{figure*}
\subsection{Scanning Electron Microscopy}
SEM images of the films are shown in Fig. \ref{fig-sem} at incremental regions from the centre towards the edge of 1" Si wafer. These images show clear radial variations in growth rate depending on the height of the sample puck. Starting with the film grown at a height of 5 mm, at a radial position of 0 mm (centre of the film) shows a large size distribution with grains as large as $\sim$\SI{1}{\micro\metre} down to the nanoscale. At 2.5 mm, these large micron size grains are still apparent, however much are fewer in distribution and a larger fraction of nanoscale grains is found. Moving further outwards, the micron size grains begin to disappear, and only nanoscale grain texturing is observed. Finally reaching the edges of the film, the grain size suddenly increases showing significant growth of a microcrystalline film. Next, for the film grown at a height of 10 mm, there is minimal radial grain size variation from 0 up to 10 mm, showing similarly large micron size grains as the 5 mm sample except has coalesced with less nanoscale grains.  However, at the edge of the film there is a clear jump again in growth rate with evidence of much larger grains. Finally, for the film grown at a height of 15 mm, the centre of the film also shows minimal radial variation from 0 up to 10 mm with the exception of a much smaller average grain size. In a similar fashion to the previous samples, the growth rate towards the edges of the film is much faster and much larger grains are found.

\subsection{Raman Spectroscopy}
\begin{figure}[t!]
  \centering
  \includegraphics[width=\figWHH\textwidth]{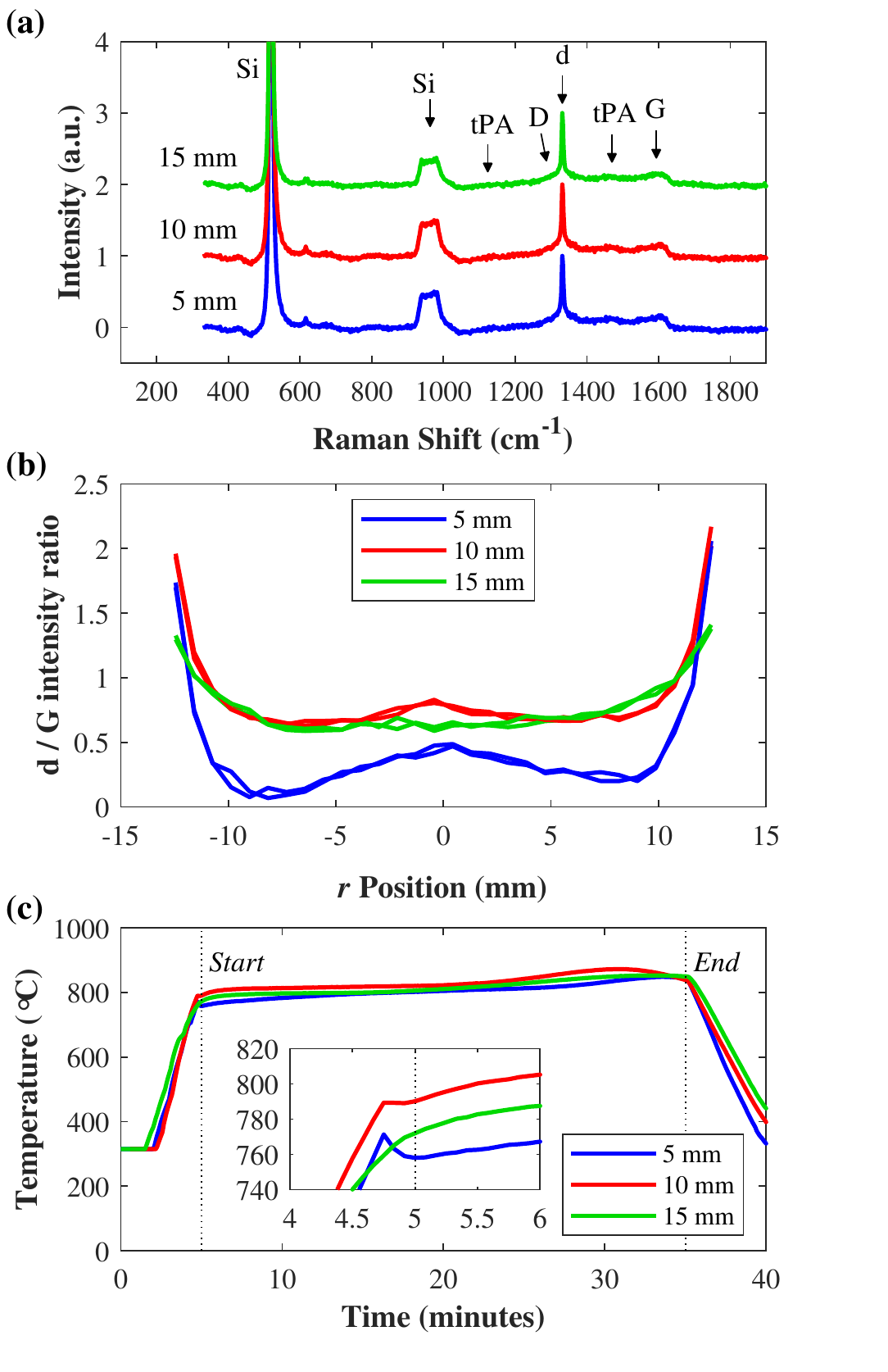}
  \caption{Raman spectroscopy of CVD diamond on 1'' Si wafers grown at holder heights of 5 mm, 10 mm and 15 mm for 30 minutes (5 kW 160 mbar). (a) shows the spectra at the centre of the wafer. The labels `d', `D', `G' and `tPA' denote the contributions from the \dia{} carbon peak in diamond, the D and G bands of \spt{} carbon and trans-polyacetylene, respectively. (b) shows a line scan of the intensity ratio of 'd' / 'G', or the implied \dia{} / \spt{} or non-diamond carbon ratio. \JACB{(c) shows the temperature recorded by the pyrometer, where the inset shows a zoom in at the start point. Pyrometer lower limit is 315 $^\circ$C.}}
  \label{fig-ram}
\end{figure}

Raman spectra at the centre of the wafers of the thin diamond films grown at different Mo puck heights are shown in Fig. \ref{fig-ram} (a). The high intensity peak at \SI{520}{\per\centi\metre} and the band at approximately \SI{950}{\per\centi\metre} are attributed to the first and second order bands of Si\cite{Prawer2004,Sedov2019}. Since the growth time is fairly short, the diamond films are fairly thin, therefore the contribution of the Si substrate is large. The sharp peak at \SI{1332}{\per\centi\metre} is attributed to the first order \dia{} carbon peak, a signature characteristic of diamond\cite{Ramaswamy1930,Bhagavantam1930,Prawer2004,Knight1989a,Ayres2017}. Amongst this diamond peak are several broad bands associated with various non-diamond carbon impurities. The weak band at approximately \SI{1420}{\per\centi\metre} and even weaker band at approximately \SI{1120}{\per\centi\metre} are both attributed to trans-polyacetylene (tPA), commonly found in CVD diamond Raman spectra although are only dominant at particularly low grain sizes such as nanocrystalline diamond (NCD)\cite{Sankaran2012a,Sankaran2018,Ferrari2000}. The broad band at approximately 1310 to 1340 cm$^{-1}$ is attributed to the $A_{1g}$ breathing mode of aromatic \spt{} carbon rings, while the peak at around 1580 to 1610 cm$^{-1}$  is attributed to the $E_{2g}$ bond stretching mode in \spt{} carbon\cite{Ferrari2000}. The low intensity of these non-diamond carbon signatures compared to the \dia{} peak at a laser excitation wavelength of 532 nm implies a low non-diamond carbon impurity concentration; in heavily \spt{} incorporated films, these band often dominate the \dia{} peak\cite{Cuenca2019,Williams2011}. 

The line scans of the d / G ratio in Fig. \ref{fig-ram} (b) show that there is a significant variation in the diamond growth across the 1" Si wafer. For all samples, the \dia{}/\spt{} peak ratio is much higher at the edges of the sample compared to the centre. This implies a much faster growth rate at the edges of the sample which is corroborated in the plasma model by the focusing effect at the samples edges, and therefore an increased plasma and reaction density. At a radial position of approximately \SI{10}{\milli\metre} from the centre, the \dia{}/\spt{} peak ratio rapidly decreases for all sample heights. For the film grown at a height of 5 mm, almost no diamond peak is found whereas for the films grown at heights of 10 and 15 mm, the \dia{}/\spt{} ratio is fairly similar towards the centre of the film. At the centre, a noticeable hump in the \dia{}/\spt{} ratio is found, which gradually disappears with increasing puck height. This variation is again caused by the variation in the plasma density across the sample; as shown in Fig. \ref{fig-model} the plasma tends towards the classic central ellipsoidal shape at shallow heights and pushes towards the edges for taller pucks at high MWPD.

\begin{figure}[t!]
  \centering
  \includegraphics[width=\figWH\textwidth]{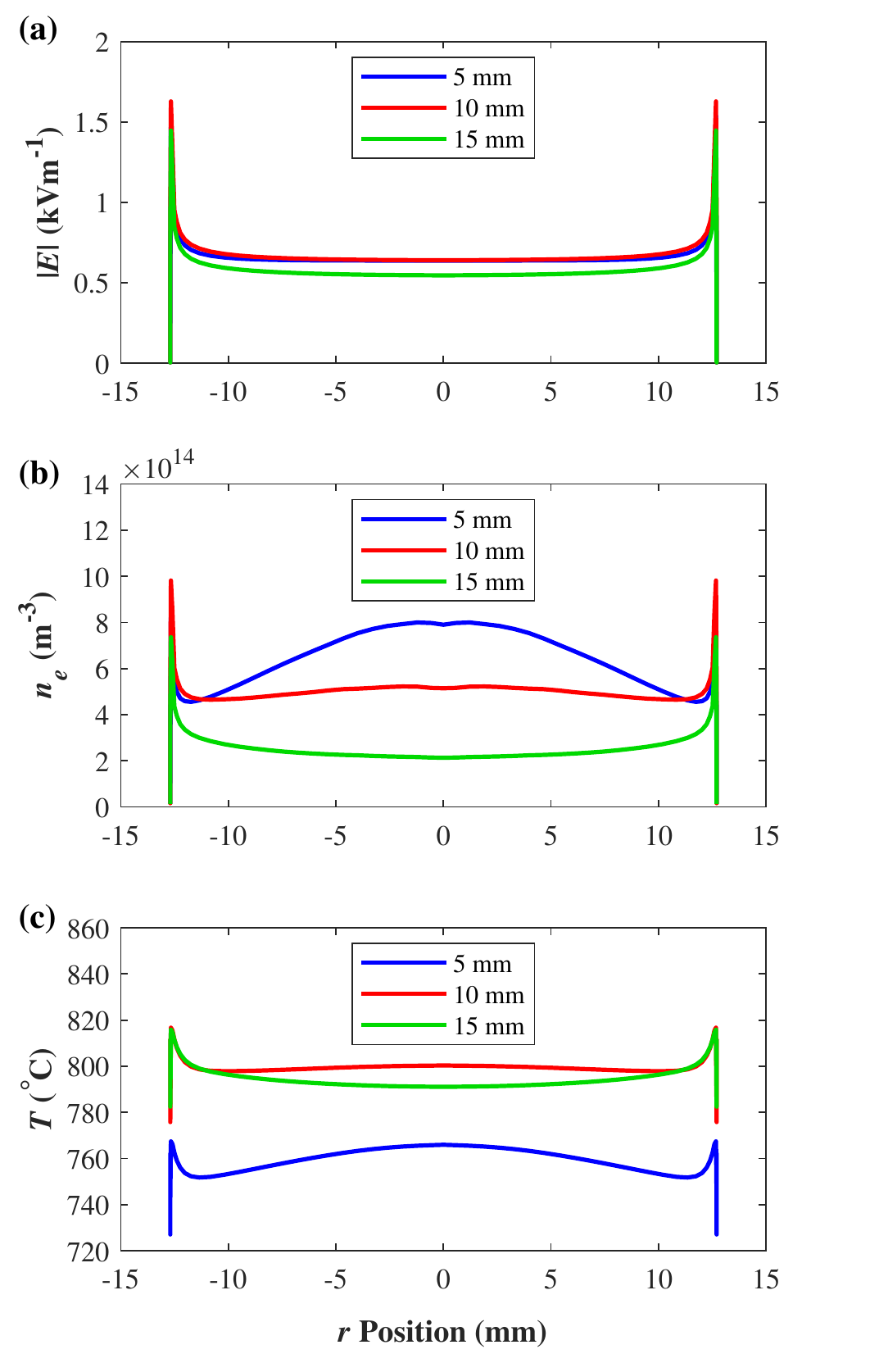}
  \caption{Model line scans over the Si sample surface of (a) the E-field magnitude in the EM model, (b) the plasma electron density and (c) the substrate temperature.}
    \label{fig-model-lines}
\end{figure}

\section{Discussion}
It is clear from the modelling and experimental results that the Mo sample puck height has a considerable effect on the plasma shape and therefore the spatial diamond growth rate across a 1" Si wafer. For comparison between the model and the experiment, Fig. \ref {fig-model-lines} shows line scans of the E-field magnitude, plasma electron density and substrate temperature across the sample from the EM, plasma and heat transfer solutions, respectively. Clear correlations and limitations in the model are identified when compared with the SEM and Raman data. For the EM model, the E-field distribution only shows the high intensity regions at the edges of the sample and almost no sign of the broad hump in the centre. This highlights a limitation of the E-field modelling approach in predicting growth rate variations across small samples. However, the plasma fluid model shows that there is a clear bump at the centre of the film for the 5 mm puck height and then becomes increasingly flatter as the puck height is increased. The electron density alone however implies that the growth rate would be much faster for a shallower puck over a taller puck which does not appear to not be the case from the Raman line scan spectra and the SEM images. The reason is likely due to the temperature at the sample surface during growth. The experimentally measured temperature is shown in Fig. \ref{fig-ram} (c) where the temperature is marginally lower for the 5 mm puck (\SI{\sim760}{\celsius}) when compared to the 10 mm (\SI{\sim790}{\celsius}) and 15 mm (\SI{\sim780}{\celsius}) pucks. In fact, the films with the most consistent quality are those grown on the 10 mm puck, with growth temperatures closer to the typical CVD diamond growth temperature of $\sim$\SI{800}{\celsius}. Thus, the electron density plasma model is only representative of some spatial growth variation but less accurate when used to compare between holders. The heat transfer solution provides a better insight as is shown in Fig. \ref{fig-model-lines}, where the temperature line scans show both the radial variation in temperature and that the 10 mm puck has a higher growth temperature than the others, thereby producing a better quality film. 

The result from this experiment is that shallower pucks are heavily cooled by the stage which reduces the substrate temperature for a given process growth condition. As the height of the puck increases, the sample is pushed into the plasma and the substrate temperature increases. However, if the puck is too tall, the plasma is largely perturbed and then focuses to the edges and away from the sample substrate. Additionally, since a larger volume of Mo is in contact with the cooled stage, the $T_g$ is reduced and therefore the substrate temperature. This model demonstrates that if the users goal is to optimise spatial homogeneity across a sample then this puts precedence on monitoring the spatial temperature across the sample during growth. 

\section{Conclusion}
Modelling of microwave hydrogen plasmas can offer simple and cost effective insights into how sample holder designs can affect MPCVD processes. This work demonstrates that EM eigenfrequency models show how sample holder pucks of different width and diameter can perturb the reactor frequency where in this instance, taller pucks have a much larger effect on frequency.  Additionally, strict EM modelling approaches have some limitations at both low and high MWPD and are generally useful for identifying possible regions for the plasma to spark, but do not necessarily describe the plasma shape. Fully coupled EM/plasma models offer a better description which models the power and pressure size and electron density. Using multi-physics coupling of EM, plasma and heat transfer solutions, the spatial variation in diamond growth can be estimated through variations in the substrate temperature. 

\section{Acknowledgements}
This project has been supported by Engineering and Physical Sciences Research Council (EPSRC) under the GaN-DaME program grant (EP/P00945X/1) and the European Research Council (ERC) Consolidator Grant under the SUPERNEMS Project (647471). SEM was carried out in the cleanroom of the ERDF-funded Institute for Compound Semiconductors (ICS) at Cardiff University.

\bibliographystyle{elsarticle-num-nourl.bst}
\bibliography{\jobname}

\end{document}